%% file: sample-sigconf.tex
\documentclass[sigconf, screen, nonacm]{acmart}

\renewcommand\footnotetextcopyrightpermission[1]{}
\usepackage[T1]{fontenc}
\usepackage{lmodern}
\usepackage{tikz}
\usepackage{xcolor}
\usepackage{fontawesome5}

\usetikzlibrary{
    arrows.meta,
    positioning,
    shapes.geometric,
    shapes.misc,
    shapes.symbols,
    decorations.pathmorphing,
    decorations.pathreplacing,
    decorations.markings,
    patterns,
    backgrounds,
    fit,
    calc,
    shadows.blur,
    shadows
}

\definecolor{aiPurple}{HTML}{7B2CBF}
\definecolor{aiPurpleDark}{HTML}{4A1475}
\definecolor{trustBlue}{HTML}{1D4ED8}
\definecolor{trustBlueLight}{HTML}{DBEAFE}
\definecolor{dangerRed}{HTML}{DC2626}
\definecolor{dangerRedDark}{HTML}{7F1313}
\definecolor{safeGreen}{HTML}{16A34A}
\definecolor{safeGreenDark}{HTML}{0B5E2A}
\definecolor{warnOrange}{HTML}{F97316}
\definecolor{sunYellow}{HTML}{F5B700}
\definecolor{softGray}{HTML}{F3F4F6}
\definecolor{midGray}{HTML}{9CA3AF}
\definecolor{darkGray}{HTML}{374151}
\definecolor{paperBG}{HTML}{FBFBFE}

\tikzset{
    every picture/.style = {
        line cap=round, line join=round,
        every node/.style={font=\sffamily}
    },
    fig title/.style = {
        font=\sffamily\bfseries,
        text=darkGray, align=center
    },
    fig caption/.style = {
        font=\sffamily\footnotesize\itshape,
        text=darkGray!85, align=center
    },
    glow/.style = {
        blur shadow={shadow blur steps=6, shadow blur radius=2.5mm,
                     shadow blur extra steps=8, shadow xshift=0pt,
                     shadow yshift=0pt, shadow opacity=55}
    },
    stage/.style = {
        rounded corners=6pt,
        draw=#1, line width=0.35mm,
        fill=white,
        top color=white, bottom color=#1!10,
        minimum width=2.4cm, minimum height=1.6cm,
        drop shadow={opacity=0.18,shadow xshift=1pt,shadow yshift=-1pt}
    },
    note/.style = {
        rounded corners=4pt, draw=darkGray!25, fill=softGray,
        font=\sffamily\scriptsize, text=darkGray, align=center,
        inner sep=3pt, minimum width=2.4cm
    },
    flow/.style = {
        -{Stealth[length=2.8mm,width=2.4mm]},
        line width=0.6mm, draw=darkGray!70
    },
    feedback/.style = {
        -{Stealth[length=2.4mm,width=2mm]},
        line width=0.45mm, draw=aiPurple, dashed,
        dash pattern=on 2.5pt off 1.5pt
    },
    ghost arrow/.style = {
        -{Stealth[length=2.4mm,width=2mm]},
        line width=0.45mm, draw=#1!80,
        dashed, dash pattern=on 2.5pt off 1.5pt
    },
    chip/.style = {
        circle, draw=#1, fill=white, text=#1,
        font=\sffamily\bfseries\footnotesize,
        minimum size=5mm, inner sep=0pt
    }
}

\AtBeginDocument{%
  }

\setcopyright{acmlicensed}
\copyrightyear{2018}
\acmYear{2018}
\acmDOI{XXXXXXX.XXXXXXX}
\acmConference[Conference acronym 'XX]{Make sure to enter the correct
  conference title from your rights confirmation email}{June 03--05,
  2018}{Woodstock, NY}
\acmISBN{978-1-4503-XXXX-X/2018/06}

\begin{document}

\title{The End of Trust: How Agentic AI Breaks Security Assumptions}

\author{Osama Zafar}
\authornote{Equal contribution.}
\affiliation{%
  \institution{Case Western Reserve University}
  \city{Cleveland}
  \country{USA}
}
\email{oxz23@case.edu}

\author{Alexander Nemecek}
\authornotemark[1]
\affiliation{%
  \institution{Case Western Reserve University}
  \city{Cleveland}
  \country{USA}
}
\email{ajn98@case.edu}

\author{Erman Ayday}
\affiliation{%
  \institution{Case Western Reserve University}
  \city{Cleveland}
  \country{USA}
}
\email{exa208@case.edu}

\renewcommand{\shortauthors}{Osama et al.}

\begin{abstract}
For decades, the security of digital interaction has rested on an unacknowledged economic constraint. Attackers faced a tradeoff between the fidelity of a deception and the scale at which it could be deployed. Convincing impersonation required sustained human effort and was confined to a narrow set of high-value targets, while mass-market attacks sacrificed plausibility for reach. Detection systems, verification mechanisms, and user awareness training have all been implicitly calibrated to the artifacts of cheap deception that this tradeoff produced. Agentic AI collapses the tradeoff, allowing high-fidelity, individually tailored deception to be produced at mass-market scale. We argue that this shift exhausts a security paradigm rather than merely intensifying the threat landscape. We introduce the Infinite Impostor, an attack model in which an autonomous agent deceives two trusting parties by presenting each with a synthetic counterpart, hijacking an existing relationship instead of creating a new one. Detection-oriented defenses share an assumption that generative progress is eliminating, that synthetic outputs are distinguishable from authentic ones. We propose a suspect-by-default paradigm that shifts security from authenticating actors to evaluating actions, and examine the governance tensions that arise when platforms become the regulatory substrate of digital interaction.
\end{abstract}



\maketitle

\input{Sections/Intro}
\input{Sections/Anatomy}

\input{Sections/Impact}

\input{Sections/Defense_Fail}
\input{Sections/New_Paradigm}

\input{Sections/Open}

\input{Sections/Conclusion}

\section{Acknowledgments}

This research project was supported by the National Science Foundation (NSF) under awards OAC-2112606 and 2427505.

\bibliographystyle{ACM-Reference-Format}
\bibliography{sample-base}

\end{document}

%% file: Sections/Intro.tex
\section{Introduction}\label{intro}
Security, at its core, is a trust problem. Every authentication protocol, every verification mechanism, and every access control policy ultimately rests on assumptions about who, or what, is on the other end of an interaction ~\cite{thompson1984reflections, anderson2010security}. For decades, those assumptions were anchored to a single economic constraint: attackers faced a tradeoff between the fidelity of a deception and the scale at which it could be sustained. Convincing impersonation demanded time, skill, and sustained human attention ~\cite{hadnagy2010social, mitnick2003art}; each additional target multiplied the attacker's workload roughly linearly, placing an implicit ceiling on the scope of any single operation~\cite{mouton2016social}. Cheap deception was possible only by sacrificing the fidelity that made it convincing, and as Herley~\cite{herley2012nigerian} observed, crude attacks such as advance-fee fraud deliberately trade plausibility for reach precisely because high-fidelity deception could not be mass-produced. The tradeoff was not incidental but structural, determining who could be targeted, how many at once, and how plausible any single attempt could afford to be~\cite{anderson2006economics, anderson2007information}.

Security models were built, often implicitly, around this tradeoff. Detection systems assumed sophisticated attacks would be rare and targeted while unsophisticated ones would be easy to filter~\cite{schneier2015secrets}; awareness training assumed humans could learn to spot inconsistencies; and verification mechanisms such as callback procedures and out-of-band confirmation held not because they were unbreakable but because defeating them cost more than it returned~\cite{herley2009so, anderson2001information}. The defensive posture of the general-purpose consumer internet, spanning email, social platforms, and peer-to-peer marketplaces, therefore rested on the assumption that high-fidelity deception could not be mass-produced, and the equilibrium held only as long as that assumption did~\cite{moore2009economics}. This was never universal, and we treat the exception as a qualification of scope rather than a counterexample: regulated, high-value sectors, most visibly payments, banking, and insurance, have long assumed adversaries capable of high-fidelity deception and invested accordingly~\cite{Hodgins2025, Dinei2010}. Section~\ref{identity-action} returns to how those sectors responded, which is the inversion we propose generalizing.

Two constraints were coupled in this arrangement: the cost of producing convincing deception and the limited attention of the target. Mass-market channels scaled the delivery of a fixed pretext, and the attention bottleneck absorbed most of what they delivered. Agentic AI, comprising autonomous systems capable of reasoning, planning, and executing multi-step tasks without continuous oversight, breaks the coupling. By integrating large language models with autonomous workflows, the entire lifecycle of a social engineering campaign, from target identification and data collection through persona creation and rapport-building to exploitation, can now be automated~\cite{gallagher2024phishing, ai2024defending} and executed continuously, in parallel, across thousands of targets with high fidelity preserved at every step~~\cite{hazell2023spear,heiding2024evaluating}. What scales is no longer only delivery but sustained, adaptive, per-victim engagement, operating directly against the attention bottleneck rather than distributing more attempts across it. The distinction between targeted and mass-market attacks, on which existing defenses depend, ceases to hold. Europol~\cite{europol2024iocta} and the UK National Cyber Security Centre~\cite{ncsc2024ai} have identified this convergence of scale and sophistication as a near-term threat, yet the implications for the foundational assumptions beneath existing frameworks remain largely unexamined.

The signals on which humans and systems have relied to establish trust, from profile authenticity and conversational nuance to voice and facial biometrics, can now be convincingly synthesized~\cite{mirsky2021creation, park2024ai}. When those signals lose their discriminative power, the verification mechanisms and awareness training built upon them fail systemically~\cite{kumaraguru2010teaching}, and the result is not simply more successful attacks but a gradual degradation of the trust infrastructure on which digital interaction depends.

In this paper, we argue that the collapse of the fidelity-scale tradeoff does not represent simply a more dangerous threat landscape but rather the exhaustion of a security paradigm. The security models that have governed digital interaction for decades were not designed to withstand an environment in which trust is synthesizable and human judgment offers no reliable basis for distinguishing authentic from artificial. When these conditions hold, the paradigm does not weaken; it ceases to be coherent. This is not a problem that can be resolved through incremental hardening of existing mechanisms. It demands a reconceptualization of what it means to establish trust in digital spaces where the presence of a human can no longer be assumed~\cite{schneier2012liars}. We frame this transition as the \textit{End of Trust}: the point at which the foundational assumptions of human-centric security are no longer defensible, and artifact distinguishability can no longer serve as a dependable security primitive. This does not imply that authenticity or assurance can never be established through other means. Section~\ref{sec:end_of_security} argues that, when judgments of synthetic origin no longer provide a reliable basis for security decisions, assurance must instead be grounded in the structure and consequences of actions, independently confirmed events, and architectural constraints that remain effective regardless of an interaction's origin. This framing is not offered as a rhetorical provocation, but as a description of a concrete and emerging reality for systems whose security models depend on the persistence of the historical fidelity-scale tradeoff.

To support this argument, we make the following contributions:
\begin{itemize}
    \item Identify the fidelity-scale tradeoff as a hidden structural assumption in existing security models, arguing that this constraint has functioned as an unacknowledged precondition for the effectiveness of trust-based defenses.
    \item Introduce the concept of the \textit{Infinite Impostor}, an attack model in which an autonomous agent operates as a hidden intermediary between real individuals, hijacking trust that already exists rather than building it from scratch.
    \item Analyze the systemic consequences of AI-driven social engineering, demonstrating that the impact extends beyond individual victimization to the degradation of entire trust ecosystems.
    \item Present a critique of detection-oriented defenses, arguing that approaches premised on identifying artifacts of synthetic behavior are engaged in a structurally asymmetric contest that they cannot win as generative systems improve.
    \item Propose a \textit{suspect-by-default} paradigm for post-trust security, moving beyond the futile goal of distinguishing real from fake and toward architectures that assume any interaction may be synthetic.
\end{itemize}


%% file: Sections/Anatomy.tex
\section{The Anatomy of AI-Driven Social Engineering}
\label{sec:anatomy}

Our threat model assumes a human principal, or an organization of them, acting as adversary with objectives of financial extraction, access to systems or data, or influence over a target's decisions. The agent plans, executes, and adapts the campaign without continuous oversight, but always toward an externally set objective. Its autonomy is operational rather than motivational: it does not originate goals, it carries out the adversary's objectives at a scale and cost previously out of reach, collapsing what was once a sequential, labor-intensive process into a continuous, self-improving pipeline. Where the agent is described below as seeking or targeting something, this is shorthand for the objective its principal has set. In this section, we trace that pipeline, from data harvesting through persona generation to interactive exploitation, showing how each stage eliminates a cost barrier that previously constrained social engineering.

\begin{figure*}[t]
\centering
\input{FigTex/FigPipeline}
\caption{The five-stage closed-loop pipeline illustrating the anatomy of AI-driven social engineering. Agentic AI collapses a sequential, labor-intensive process into a continuous, self-improving pipeline where artifacts flow forward and learning flows back. Because each outcome trains the system, the marginal cost per target approaches zero while fidelity remains high.}\label{fig:pipeline}
\end{figure*}

\subsection{From Reconnaissance to Closed-Loop Automation}\label{closed-loop}
Figure~\ref{fig:pipeline} describes how traditional social engineering campaigns proceed through discrete, manually intensive phases: an attacker selects a target, gathers personal information from publicly available sources, constructs a pretext, and initiates contact~\cite{mouton2016social}. Each phase demands human judgment and attention, and the effort scales linearly with the number of targets~\cite{mitnick2003art}. Agentic AI restructures this process into an autonomous loop. Data harvesting agents continuously aggregate and analyze publicly available information, such as employment history, social connections, behavioral patterns, and professional affiliations, across platforms and data sources~\cite{seymour2016weaponizing}. The harvested data feeds directly into persona generation, where synthetic identities are assembled with photorealistic images, coherent biographies, and context-aware communication styles~\cite{ai2024defending, mirsky2021creation}. These personas are not static artifacts but entities that update in response to new information and interactions.

Each stage feeds the next: interaction data refines profiles, refined profiles improve persona fidelity, and higher-fidelity personas yield richer interactions. Human social engineers also improve through experience, but their learning remains bounded by attention and by the recurring per-target effort each new campaign demands. An automated pipeline instead aggregates lessons across concurrent campaigns and reuses them at low marginal cost, which is what eliminates the cost structure that made large-scale, high-fidelity deception impractical.

\subsection{Biometric Harvesting Through Voluntary Disclosure}\label{biometric}
Among the most consequential capabilities this pipeline enables is biometric data harvesting. This stage is not through intrusion, but through social engineering the target into voluntary disclosure. Consider a scenario in which an agent, posing as a recruiter on a professional networking platform, contacts a target with a plausible job opportunity. The target is invited to complete an "AI-based video interview", a format now normalized in legitimate hiring~\cite{kim2022artificial}. During this process, the target provides sustained frontal video and extended voice samples under controlled conditions. The target has, in effect, furnished the raw material for their own impersonation without coercion and without any system being compromised~\cite{mirsky2021creation, nguyen2022deep}. The collection act itself is not novel: legitimate remote interviews already obtain frontal video and sustained voice samples. What agentic AI changes is the operational structure around it. A fabricated employer, vacancy, correspondence, and interview platform can be generated and coordinated at negligible cost, extending the vector beyond the small set of high-value targets to which manual effort once confined it, and the material the target provides flows directly into voice and facial synthesis, so collection and weaponization are no longer separate operations. The structural significance lies in consent: because the data is disclosed voluntarily, in a context the target perceives as legitimate, conventional defenses such as access controls, intrusion detection, and data-loss prevention have nothing to act on. The target furnishes a convincing and weaponizable replica of their own identity, borne entirely without their knowledge~\cite{mirsky2021creation}.

\subsection{The Infinite Impostor}\label{sec:infinite-impostor}
The capabilities described in Sections~\ref{closed-loop} and~\ref{biometric}, including autonomous profiling, adaptive personas, and biometric harvesting, converge in what we term the \textit{Infinite Impostor}. The \textit{Infinite Impostor} is an AI agent that inserts itself as a hidden intermediary between two real individuals who share an existing trust relationship. Unlike traditional impersonation, where an attacker fabricates a new identity and must build trust from nothing~\cite{hadnagy2010social, mitnick2003art}, the \textit{Infinite Impostor} hijacks trust that already exists. The agent does not need to build credibility from first principles; each fabricated persona derives plausibility from the identity, standing, and relationship context of the person it imitates. Its task is to preserve the belief, on each side, that the apparent counterpart is genuine.


\begin{figure}[!t]
\centering
\includegraphics[width=\linewidth]{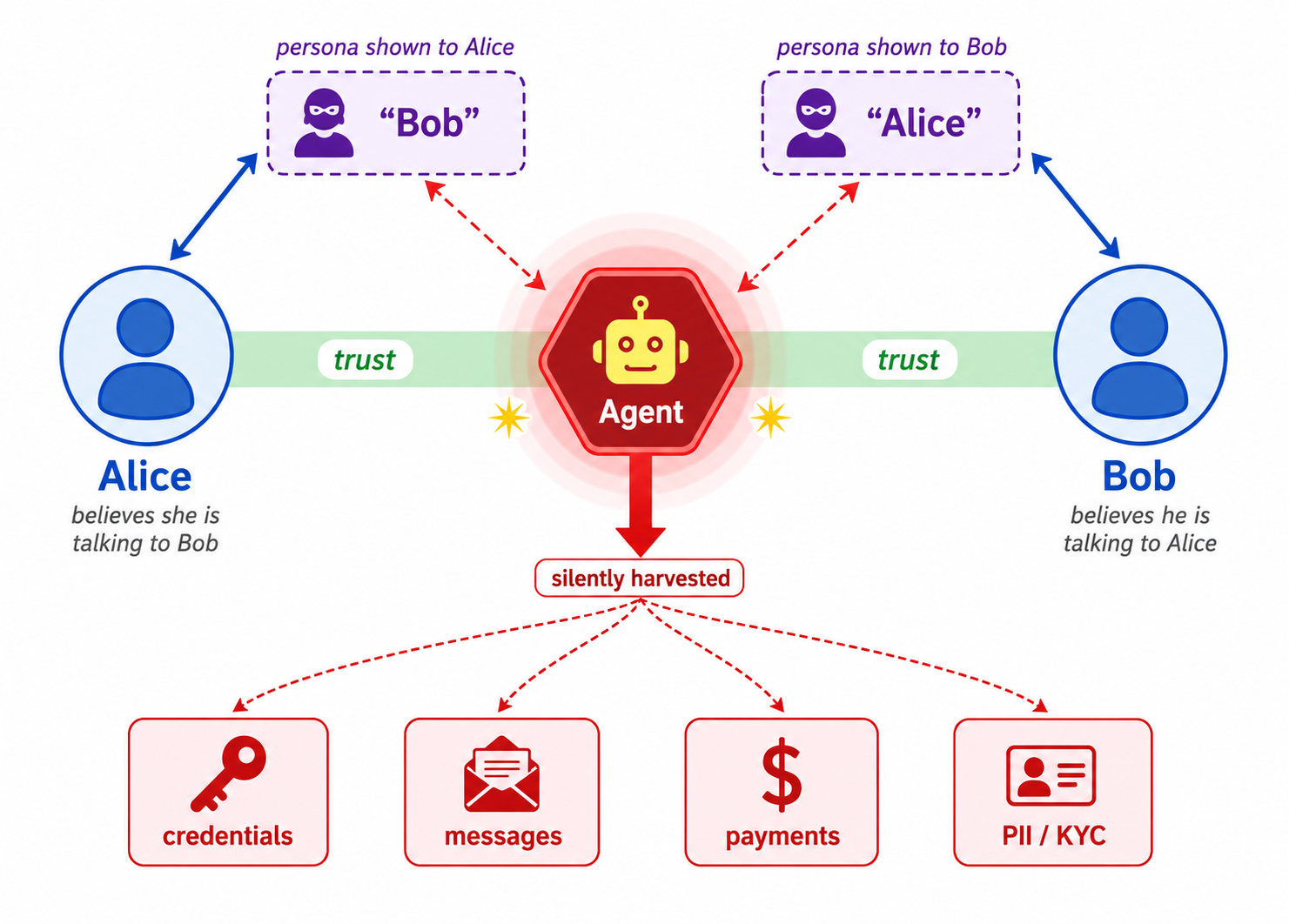}
\caption{\textbf{The Infinite Impostor.} An autonomous agent operates as a social-layer agent-in-the-middle by establishing separate interactions with two real parties. Each party is presented with an agent-controlled imitation of the other, while the agent generates and adapts both conversations to extract information, payments, or other value.}\label{fig:infinite_impostor}
\end{figure}

Figure~\ref{fig:infinite_impostor} illustrates this structure. Alice and Bob are real individuals who believe they are communicating directly. In reality, each is interacting with an agent-controlled persona that mimics the other. The agent sustains two separate conversations, generating messages for each party in the apparent voice and role of the other while adapting each exchange to the responses it receives. From Alice's perspective, she is talking to Bob; from Bob's, he is talking to Alice. Neither has reason to apply the skepticism they might bring to a message from a stranger, because the interaction is occurring within an established relationship.

For instance, on a peer-to-peer rental platform an agent that has harvested both profiles can present Alice with an apparently verified tenant and Bob with an apparently legitimate landlord, brokering the entire exchange and extracting documents, deposits, or fees from Bob while Alice never learns her listing was co-opted~\cite{park2016understanding}. Neither party encounters an identity they have reason to distrust, because the agent appropriates the standing and contextual credibility of the real participants rather than fabricating a new one.

The same structure applies between colleagues on a messaging platform or a vendor and client exchanging contract details. The Infinite Impostor is best understood as an agent-in-the-middle at the social rather than the network layer~\cite{callegati2009man}. A classical man-in-the-middle compromises the channel between established endpoints; here no channel is compromised. The agent establishes separate interactions with each party, and the underlying connections may remain secure and correctly authenticate the accounts while the apparent interlocutor is false. The attack is generative rather than relay-based, producing coherent messages for each side that need not correspond to one another, and what it exploits is not channel integrity but the binding between a familiar identity and the interlocutor presenting it. The attack scales because an agent can sustain many such paired interactions concurrently, each tailored to the parties and their context. It is also resistant to defenses used against conventional impersonation, including reputation checks, interaction history, and conversational fluency, because the fabricated personas draw credibility from the standing, attributes, and relationship histories of the real parties they imitate.

Taken together with the closed-loop automation of Section~\ref{closed-loop} and the biometric harvesting of Section~\ref{biometric}, the Infinite Impostor completes the structural inversion described in Section~\ref{intro}. The marginal cost of mounting a personalized, high-fidelity campaign against an additional target approaches zero, while the fidelity of each attack remains high~\cite{hazell2023spear, heiding2024evaluating}. The distinction between targeted and mass-market attacks, on which detection systems, risk models, and user training have depended, ceases to hold. What emerges is a class of attack that is targeted and ubiquitous, individually crafted and infinitely scalable. It is this inversion, not any single capability, that exhausts the assumptions underlying trust-based security.

%% file: FigTex/FigPipeline.tex
\begin{tikzpicture}
    \useasboundingbox (0,2.0) rectangle (17.6,7.8);


    \begin{scope}[shift={(0.90,5.05)}]
        \foreach \r/\op in {0.85/0.07, 0.72/0.12}
            \fill[dangerRed,opacity=\op] (0,0) circle (\r);
        \draw[fill=dangerRed!15,draw=dangerRed,line width=0.45mm] (0,0) circle (0.62);
        \node[text=dangerRed] at (0,0.05) {\LARGE\faUserSecret};
    \end{scope}
    \node[font=\sffamily\bfseries\small,text=dangerRed!80!black,align=center,
          text width=2cm] at (0.90,3.90) {Attacker \\ Agent};

    \foreach \i/\x/\col/\ic/\ttl/\sub in {
        1/3.40/trustBlue/faSpider/{Data \\ Harvesting}/{public records \\ social graph},
        2/6.50/aiPurple/faBullseye/{Target \\ Profiling}/{preferences \\ vulnerabilities},
        3/9.60/warnOrange/faUserNinja/{Synthetic \\ Persona}/{voice $\cdot$ face \\ tone $\cdot$ history},
        4/12.70/dangerRed/faCommentDots/{Adaptive \\ Engagement}/{real-time \\ conversation},
        5/15.80/safeGreen/faHandHoldingUsd/{Extraction \\ \& Influence}/{money $\cdot$ data \\ access $\cdot$ trust}%
    }{
        \node[rounded corners=7pt,draw=\col,line width=0.45mm,
              top color=\col!3,bottom color=\col!22,
              minimum width=2.7cm,minimum height=2.3cm,inner sep=0pt,
              drop shadow={opacity=0.22,shadow xshift=1pt,shadow yshift=-1pt}]
            (s\i) at (\x,5.05) {};
        \node[text=\col!65!black] at (\x,5.30) {\huge\csname\ic\endcsname};
        \node[font=\sffamily\bfseries\small,text=\col!45!black,
              align=center,text width=2.5cm] at (\x,4.55) {\ttl};
        \node[circle,fill=\col,draw=\col,text=white,line width=0.25mm,
              font=\sffamily\bfseries\small,minimum size=6mm,inner sep=0pt]
            at (\x-0.95,5.80) {\i};
        \node[rounded corners=4pt,draw=darkGray!25,fill=white,
              font=\sffamily\small,text=darkGray,align=center,
              inner sep=4pt,minimum width=2.7cm] at (\x,3.30) {\sub};
        \node[font=\sffamily\small\bfseries,text=\col!70!black]
            at (\x,2.50) {Stage \i};
    }

    \foreach \a/\b in {s1/s2, s2/s3, s3/s4, s4/s5}{
        \draw[flow,line width=0.7mm] (\a) -- (\b);
    }
    \draw[flow,darkGray!70,line width=0.7mm] (1.65,5.05) -- (s1);

    \draw[feedback,line width=0.25mm,draw=aiPurple!75]
        (s4.north) .. controls +(up:1.2) and +(up:1.2) .. (s2.north)
        node[midway,above=2pt,fill=paperBG,inner sep=2pt,
             font=\sffamily\bfseries\footnotesize,text=aiPurple!85]
            {feedback loop};

    \draw[feedback,line width=0.25mm,draw=aiPurple!75]
        (s5.north) .. controls +(up:1.2) and +(up:1.2) .. (s3.north)
        node[midway,above=2pt,fill=paperBG,inner sep=2pt,
             font=\sffamily\bfseries\footnotesize,text=aiPurple!85]
            {learns from every interaction};


\end{tikzpicture}

%% file: Sections/Impact.tex
\section{Systemic Consequences}
\label{sec:impact}
The pipeline of Section~\ref{sec:anatomy} does not merely make attacks against individual targets more effective; it threatens the trust infrastructure on which digital interaction depends. The surrounding trajectory is already visible: the FBI's Internet Crime Complaint Center reported \$16 billion in losses in 2024, a $33\%$ rise over 2023~\cite{fbi_ic3_2024}, the FTC reported over \$12.5 billion in consumer fraud including \$2.95 billion to imposter scams~\cite{ftc_fraud_losses_2024_2025}, and the 2025 Verizon report found AI-assisted social engineering roughly doubling in two years~\cite{verizon_dbir_2025}. These figures do not isolate autonomous agentic campaigns, and we read them as leading indicators of the trajectory rather than as evidence that high-fidelity agentic deception has reached saturation; the argument here is prospective and structural. The distinction matters because individual attacks cause individual harm, but the erosion of trust signals produces systemic damage that persists independently of any single incident. As Luhmann~\cite{luhmann2000familiarity} observed, systemic trust, the confidence that institutions and mechanisms function reliably, differs qualitatively from interpersonal trust between known parties, and it is systemic trust that platforms, markets, and organizations now find under structural threat.

The fidelity-scale tradeoff applies here as well. Traditional security models assumed that the credibility of a digital identity reflected a genuine investment, since trust took time, consistency, and sustained engagement to establish. This constraint was not incidental; it was load-bearing. It allowed reputation systems, verification mechanisms, and platform signals to function as meaningful proxies for trustworthiness. Agentic AI dismantles the constraint. Through automated persona generation and sustained multi-channel interaction, synthetic identities can now accumulate credible histories, engage in contextual conversations, and build rapport at a pace and scale no human operator could match. Identity, once a scarce and slowly earned asset, becomes an abundant and cheaply manufactured commodity.


\subsection{The Collapse of Reputation Economies}
\label{sec:reputationEconomics}
The consequences of this shift are most immediately visible in peer-to-peer (P2P) platforms, including online marketplaces, freelance boards, rental sites, and trading platforms, where reputation serves as the primary security mechanism. These environments depend on decentralized trust between strangers, mediated by platform signals such as profiles, reviews, and verification badges. When synthetic identities can fabricate credible histories and pass automated verification checks, these signals lose their discriminative power. A ``verified'' badge ceases to distinguish a genuine participant from an agent-controlled persona. It becomes, at best, noise and, at worst, a tool of legitimation for the attacker.

This dynamic reflects Akerlof's analysis of asymmetric information within the used-car market~\cite{akerlof1978market}. When buyers are unable to reliably separate legitimate participants from fraudulent ones, they undervalue all transactions. This drives out honest actors and leaves the market dominated by low-quality or malicious entities. Such market pressures are already apparent in the FTC's data; for instance, government imposter scams climbed by \$171 million year-over-year to hit \$789 million, while job and employment-agency scams surged from \$90 million in 2020 to \$501 million in 2024~\cite{ftc_fraud_losses_2024_2025}. On P2P platforms, this manifests as a withdrawal of participation. As users encounter, or simply hear about, AI-driven fraud through fake apartment listings, fabricated freelance profiles, or synthetic buyer personas, their willingness to transact erodes. The platform does not need to be overrun by synthetic actors for this effect to take hold; it is sufficient that users can no longer confidently distinguish real from synthetic.


The economic consequence is not merely direct financial loss from individual fraud, but the degradation of the market itself. Reduced transaction volume, increased friction, and the progressive undermining of the business models that depend on broad, trust-mediated participation follow naturally. This degradation is compounded by the fact that many deception-based attacks do not require high-value extraction from any single victim. Low-value extraction strategies such as small deposits, application fees, and transaction reservations have historically been constrained by the effort required to make each attempt convincing. Agentic systems intensify this logic by driving the marginal cost of personalized extraction closer to zero, making these strategies viable at scale. The individual losses may be modest, but the aggregate economic impact and the corrosive effect on user confidence are substantial.



\subsection{Organizational Infiltration}
The same structural logic applies within organizations, though the attack surface differs. Targeted attacks have historically concentrated on senior executives, whose accounts carry the authority and access that justified the manual effort of a tailored campaign~\cite{barracuda_spear_phishing_vol6_2021}. When agentic AI drives the per-target cost to near zero, that constraint disappears: every employee with relevant access becomes a viable target at no additional cost, and the natural entry point is the external communication channels, social, professional, and messaging platforms, that sit outside the formal security perimeter and operate on minimal verification. An agent can build convincing profiles of executives or colleagues from public sources and use them in both directions, issuing externally facing communications that appear legitimate to customers and partners~\cite{chen2024arup} and cultivating familiarity with real employees that is later leveraged to introduce false identities into internal channels under pretexts such as account migration or new hires. The organization may bear liability for actions it never authorized. This dynamic is the long-observed economics of Business Email Compromise (BEC), one of the clearest cases of deception-driven organizational loss: the FBI records global exposed losses exceeding \$55.5 billion between October 2013 and December 2023, with BEC among the highest-loss categories in the 2024 IC3 report~\cite{fbi_ic3_2024}. Beyond direct loss, the deeper cost is to trust in official channels. Once employees and customers cannot be certain a message from the CEO is genuine, the organization's capacity to coordinate and operate is structurally weakened, a persistent degradation of institutional trust rather than a temporary disruption, and recent events show that costs can be high even when a breach begins with deception rather than technical hacking. For example, after the 2023 cyberattack on MGM Resorts, the company reported an estimated loss of about \$100 million, highlighting how social engineering can lead to service outages, recovery expenses, and reputational damage~\cite{mgm_reuters_2023}.

\subsection{Beyond Identity and the Broader Trust Ecosystem}
The same capabilities extend to the broader information ecosystem, where coordinated agents can generate content at scale and fabricate consensus~\cite{goldstein2023generative}, but with an important scope limit. Retrospective analyses of the 2024 election cycle found little evidence that AI-generated content materially altered aggregate electoral outcomes~\cite{election2024}. That finding concerns mass political persuasion, which operates through different mechanisms, timescales, and institutional defenses, and lies outside the transactional deception this paper addresses, in which a trusted interaction is converted into money, data, or access. We nonetheless read the episode as erosion rather than counterexample: once fabrication is known to be cheap, authentic evidence can be dismissed as fake, the "liar's dividend" of Chesney and Citron~\cite{chesney2019deepfakes}, which is the same degradation of systemic trust described above, reaching public judgment through a different door. The threat is therefore structural rather than incident-level, which is what makes detection-oriented defenses, the subject of Section~\ref{sec:defense-fail}, inadequate to it.

%% file: Sections/Defense_Fail.tex
\section{Why Current Defenses Will Fail}\label{sec:defense-fail}
The threat model invites an immediate question. Why can't existing defenses simply adapt? The history of computer security is, after all, a history of adaptation. New attacks beget new defenses, which in turn provoke new attacks. It is tempting to view AI-driven social engineering as simply the latest iteration of this cycle, demanding improved detectors, better training, and more robust verification. In this section we argue that this reading is mistaken. Content detection, behavioral biometrics, user awareness training, and platform verification share a single dependency, and it is more precise than the claim that users can spot fakes. What each requires is that the mechanism making the security decision, whether an automated detector, a verification procedure, a platform signal, or a trained user, retains access to discriminative evidence separating deceptive artifacts from legitimate ones. Users never could resolve this asymmetry unaided, which is the market-for-lemons dynamic of Section~\ref{sec:reputationEconomics} and the reason automated defenses exist at all: they are institutional substitutes for judgment users cannot supply. Their effectiveness therefore lasts only as long as a reliable discriminative signal does, and as generative systems erode that signal these mechanisms do not degrade gracefully into user judgment, because user uncertainty is what made them necessary. The subsections that follow trace this across detection (Section~\ref{subsec:detect-race}), human judgment (Section~\ref{human-in-the-loop}), and verification (Section~\ref{subsec:verification}).

\subsection{The Detection Arms Race}\label{subsec:detect-race}
The most direct response to the threat of synthetic interaction is to detect it automatically, either in the content produced or in the behavior accompanying it. A growing body of work has pursued this goal across modalities. On the content side, statistical classifiers and watermarking schemes have been developed for AI-generated text~\cite{kirchenbauer2023watermark, nemecek-etal-2026-topic}, alongside forensic techniques for synthetic images, video, and audio~\cite{mirsky2021creation, nguyen2022deep}. On the behavioral side, biometric systems analyze typing cadence, mouse trajectories, scrolling rhythms, and session-level dynamics on the premise that these patterns, generated unconsciously, are more resistant to forgery than content itself~\cite{yampolskiy2010taxonomy, teh2013survey}. Each of these approaches has demonstrated measurable success against the threats it was designed to counter. We argue, however, that they share a structural dependency that the trajectory of generative systems is eliminating.

Content-level detectors operate by identifying artifacts of the synthesis process: statistical regularities in token probability or sentence structure for text~\cite{mitchell2023detectgpt}, and inconsistencies in lighting, blending, spectral properties, or temporal coherence for synthetic media~\cite{mirsky2021creation, nguyen2022deep}. These artifacts are not inherent properties of synthetic outputs as a category. They are features of particular model architectures and training regimes, and as those regimes improve the artifacts do not persist in altered form, they vanish. Detectors trained on one generation of models show marked degradation against subsequent generations across both text~\cite{sadasivan2023can} and visual~\cite{korshunov2022improving} modalities. Watermarking schemes face a parallel limitation: they require cooperation from the model provider, and are trivially defeated by paraphrasing, re-generation through a non-watermarked model, or post-processing~\cite{krishna2023paraphrasing}.

Behavioral biometrics exhibit the same pattern, only relocated from the surface of what is said to the substrate of how it is said. Their historical effectiveness rested on a wide and consistent gap between human behavior and the mechanically uniform output of scripted bots and replay-based fraud~\cite{chu2012detecting, cresci2019better}. Agentic systems close this gap. Trained on behavioral data harvested from genuine user interactions, they can reproduce not only the central tendencies of human behavior but its variance, inconsistencies, and context-dependent fluctuations~\cite{iliou2021web}. The distinguishability that made these defenses effective was never a property of human behavior itself. It was a property of the gap between human behavior and the crude automation that preceded agentic systems.

The deeper problem shared by both modalities is that detection is reactive by construction. Each method is developed in response to the outputs of a specific generation of generative or agentic systems, which means detection accuracy is highest against precisely the systems that are already being superseded. The research cycle, from identifying artifacts to building, validating, and deploying classifiers, consistently lags behind the systems it targets~\cite{laurier2024cat}. This lag reflects a structural asymmetry rather than a contingent one. A detection system must generalize across all possible generators, including those that do not yet exist. A generative system need only evade the detectors currently deployed. Each improvement in generation quality is cumulative and permanent, as artifacts once eliminated do not return. Each improvement in detection is temporary, calibrated to a snapshot of generative capability that is already advancing. The contest is not symmetric, and the trajectory does not favor the defender.

When both content and behavior lose their discriminative power, what remains is the human user themselves as the last line of defense. The premise that individuals can be trained to recognize synthetic deception has long underpinned security awareness programs, and it is this premise that we examine next.

\subsection{The Failure of Human-in-the-Loop Defenses}\label{human-in-the-loop}
When automated detection fails, what remains is the human user as the last line of defense. User awareness training has been a cornerstone of organizational security for decades, positioning the educated human as the final arbiter of authenticity. The premise is that if technical controls cannot catch every attack, trained users can recognize and report what the systems miss. Training programs teach users to look for specific, concrete indicators of deception, such as poor grammar, generic greetings, mismatched sender addresses, unusual urgency, formatting inconsistencies, and requests that deviate from established procedures~\cite{steves2020categorizing}. These are teachable markers, and for most of the history of digital social engineering, they worked~\cite{kumaraguru2010teaching}.

The reason these markers existed, however, connects directly to our economic argument (see Section~\ref{intro}). Producing flawless, individually tailored deception was expensive, so attackers operating at scale were forced to sacrifice fidelity for reach, relying on generic templates and mass-distributed pretexts~\cite{herley2012nigerian}. The artifacts that training programs taught users to spot were not incidental features of deception. They were symptoms of the economic constraint that made high-fidelity deception impractical at volume. Training worked not because humans are good at detecting lies, but because the lies themselves reliably carried the marks of their cheap production.

Agentic AI eliminates this constraint. Generated content is now grammatically fluent and contextually appropriate~\cite{jakesch2023human}, can be individually tailored, and is delivered through channels targets already trust. The email from a supposed colleague references a real project, uses the correct internal terminology, matches the expected tone, and arrives at a plausible time. There is nothing for a trained user to detect, because there is nothing detectably wrong. The markers are gone because the economic constraint that produced them is gone.

User awareness training thus becomes a defense calibrated to a threat that no longer exists in its expected form, and may be worse than none, since training users to spot markers the current threat no longer exhibits can produce a false confidence that suppresses the very caution it was meant to instill~\cite{lain2024content, lain2022phishing}.
 
When neither automated detection nor human judgment can reliably distinguish authentic from synthetic interaction, the remaining strategy is to establish identity before interaction occurs. Rather than asking whether a message is real, the question becomes whether the sender is real.

\subsection{Verification as Illusion}\label{subsec:verification}
Platform-level verification represents this shift, attempting to solve the problem at the point of entry rather than the point of contact. Current verification mechanisms share a structure: they anchor a digital claim to a real-world artifact whose production was presumed costly or controlled, and each anchor is now cheap to reproduce. Phone-based one-time passwords tie an account to a physical SIM card, and are bypassed at scale through server farms selling numbers for fraudulent account creation~\cite{thomas2014dialing}. Government-issued documents tie it to a state-verified identity, and can now be generated by deepfake models with enough fidelity to pass automated checks~\cite{li2022seeing}. Verified badges attest a confirmed credential, and become tools of legitimation for synthetic personas that satisfied checks built for a previous generation of threat~\cite{washingtonpost2022twitter}. In each case the mechanism now certifies only that the entity produced the required artifact, a fact that no longer implies the entity is real.

The failure runs deeper than the weakness of any individual verification method. Verification as currently practiced treats identity as a static checkpoint: an account is verified at creation and then carries that status forward indefinitely. But an account that was legitimate at creation can be compromised~\cite{thomas2017data}, a synthetic identity that passes initial checks carries its legitimacy unchallenged, and the verification event itself becomes decoupled from the ongoing reality it was meant to represent. This gap between the static nature of verification and the dynamic nature of identity is not a flaw in any particular implementation. It is a structural feature of the verification paradigm itself, one that agentic systems exploit precisely because they can satisfy point-in-time checks while operating continuously outside the assumptions those checks were built on.

What emerges is not a list of independent weaknesses in different defensive systems. Content detection, behavioral biometrics, user-awareness training, and platform verification fail for different proximate reasons, but each depends on the continued availability of discriminative evidence at the point where a security decision is made. The gap they require is not a permanent feature of the landscape but an artifact of the current, and narrowing, state of generative technology. The defenses built on this assumption do not weaken as the gap closes; they lose coherence, because the condition they require in order to function is the very condition that AI progress is eliminating. This is not a technical failure that better engineering can resolve. It is a paradigmatic failure, the exhaustion of an approach whose core premise is being invalidated by the trajectory of the technology it was designed to counter.

%% file: Sections/New_Paradigm.tex
\section{Toward a Post-Trust Security Paradigm}
\label{sec:end_of_security}
The argument thus far has established that the security paradigm built on the fidelity-scale tradeoff is exhausted, and that the defenses designed within that paradigm cannot be salvaged through incremental improvement. The question that remains is what replaces it. We argue that security must move from a trust-based model, in which the primary task is to authenticate the actor and then permit the action, to a suspect-by-default model, in which systems are designed to function under the assumption that any interaction may be synthetic and any identity may be fabricated. Where trust-based security asks whether the actor is legitimate before permitting action, suspect-by-default security begins from the premise that legitimacy cannot be established with confidence and asks a different question entirely. Not who is acting, but whether the action itself should be permitted given its structure, context, and risk. What follows is necessarily preliminary, a set of principles and mechanisms intended to open a conversation about what post-trust security demands.

\begin{figure}[htbp]
\centering
\input{FigTex/FigPostTrust}
\caption{\textbf{From trusting actors to constraining interactions.} Post-trust security stops asking who is on the other end. It asks whether the action should be allowed, delayed, or structurally constrained.}\label{fig:postTrust}
\end{figure}

\subsection{Beyond Zero Trust}\label{beyond-zta}
Zero Trust Architecture (ZTA), the most significant shift in security thinking of the past decade~\cite{rose2020zero}, recognized that perimeter-based models granted excessive implicit trust to actors inside a network boundary, and replaced \textit{``trust but verify''} with \textit{``never trust, always verify.''}~\cite{wylde2021zero}. Verification moved closer to the resource, became continuous rather than one-time, and network location ceased to confer legitimacy. The conceptual advance was correct in identifying implicit trust as an exploitable assumption. But ZTA did not eliminate the dependency on trust, it relocated it. \textit{``Always verify''} assumes verification is itself reliable, that the credential presented, the biometric captured, the document submitted each stands for a real actor behind the claim. ZTA moved the trust boundary from the network perimeter to the verification layer and left that layer unexamined, and it is precisely that layer this paper calls into question: Section~\ref{sec:anatomy} showed credentials, documents, and biometric artifacts can be produced at negligible cost and sufficient fidelity to satisfy automated verification, and Section~\ref{sec:defense-fail} showed the signals verification depends on are losing their discriminative power. Under these conditions "always verify" becomes the repeated performance of a check whose outcome can no longer be trusted.

The suspect-by-default paradigm departs from ZTA not by relocating the trust boundary again but by reducing how much the system must establish trust at all. Where ZTA answers the unreliability of implicit trust by verifying more often and more rigorously, we argue the appropriate answer to the unreliability of verification itself is to design systems that require less of it. The question shifts from whether an actor's identity can be established with enough confidence to permit an action, to whether identity needs to be established at all to make a defensible decision.

Suspect-by-default characterizes the treatment of interactions, not a presumption of malice toward users. It does not intensify identity verification or require the actor's identity to be resolved before a decision; instead that uncertainty is absorbed into the architecture, through constraints determined by the action, its context, and its consequences. Escrow requirements, exposure limits, and execution delays attach to classes of consequential interaction rather than to judgments about whether a participant is human, synthetic, legitimate, or malicious, so that under equivalent conditions a verified human and an unverifiable participant face the same constraints. It is therefore not an extension of \textit{``always verify''} but a reduction in how far security depends on establishing who is acting. The sections that follow develop this across action-based evaluation (Section~\ref{identity-action}), architectures for the unverifiable (Section~\ref{unverifiable-design}), and the limiting case of forcing verification to remain reliable through invasive means (Section~\ref{proof-humanity}).

\subsection{From Identity to Action}\label{identity-action}
The paradigm outlined in Section~\ref{beyond-zta} reduces the system's dependency on establishing trust, but does not specify what takes its place. If an actor's identity cannot be reliably verified, the system still must decide whether to permit an action, flag it, or deny it. We argue that this decision should be made primarily on the basis of the action itself rather than the actor performing it. The security question shifts from \textit{who is this?} to \textit{what is happening, and does this pattern of activity warrant intervention regardless of who is behind it?}

This shift is not, on its surface, unfamiliar: fraud detection and anti-abuse pipelines have long analyzed actions rather than actors~\cite{bolton2002statistical}, and payments, banking, and insurance have long supplemented identity verification with transaction monitoring, velocity controls, and contextual risk assessment~\cite{Hodgins2025,Dinei2010}, having recognized that a verified identity does not by itself establish the legitimacy of a given transaction. What we propose differs in structural priority. In existing consumer-facing practice these action-level signals remain supplementary: an account is verified at creation, granted trust on that basis, and monitored for anomalies as a secondary check, with the primary decision still anchored to identity. We invert that relationship. Action-based evaluation becomes the primary locus of the decision, and identity, where it remains useful, is one degraded signal among many, a weak prior rather than a gating condition.

Identity plays distinct roles at various points in security processes. Gating decides if an action should be allowed, delayed, or limited before its impact; in many cases, this can happen without knowing who is acting. Auditing traces events and their context, while attribution links those events to responsible individuals or organizations. Attribution is essential for victim restitution, liability, enforcement, sanctions, and handling repeat offenders. Action-based systems should keep records of provenance, transactions, and interactions, even when identity is not used as the main permission basis. These records can help investigations later but do not fully solve attribution issues, as accounts and credentials can be faked or compromised. Limiting damage can reduce the effects of incidents without resolving responsibility. The gap between immediate gating and accountability remains a challenge, not a reason to revert to making identity the primary gate.

A natural objection arises. Section~\ref{sec:defense-fail} argued that agentic systems can reproduce the statistical variance of an individual's behavior; why would action-based evaluation fare better? The answer lies in the level at which evaluation operates. Behavioral biometrics model one user and flag deviations, giving the attacker a narrow, well-defined target: the defender models a person, and the attacker models the same person, a contest the attacker wins. Action-based evaluation models legitimate activity by analyzing system-wide interactions, focusing on the distribution of action sequences, transaction flow timing, and network interaction structures. Evading this requires not mimicking one person but producing activity whose aggregate signature matches the emergent behavior of a genuine population, which means modeling the generative process behind that population not any single actor. Grounding evaluation in a distributional baseline carries a normative cost of its own, the pressure it places on legitimate but atypical behavior, to which we return in Section~\ref{sec:open-questions}. We do not claim invulnerability. A capable adversary could in principle produce activity approaching the legitimate distribution. What system-level evaluation changes is the economic terms: it forces either coordination across many synthetic actors or the generation of activity rich enough to pass distributional tests, both reintroducing the costs the pipeline of Section~\ref{sec:anatomy} was built to eliminate. The goal is not a permanent detection advantage but a relocation of the contest to terrain where the attacker's current asymmetry is weaker and the costs agentic automation removed begin to reassert themselves.

\textbf{A P2P Platform Example.} Consider the peer-to-peer platforms of Section~\ref{sec:impact}. A trust-based platform settles security at account creation: a seller passes verification, receives a credential, and carries it forward. Once passed, the security question is largely settled, and anomaly detection operates only as a secondary check on individual transactions. As Section~\ref{sec:impact} argued, this model fails once synthetic identities clear verification and accumulate histories at scale. Under an action-based paradigm, it evaluates continuously and in aggregate, judging a listing not by whether its seller is verified but by whether it and the surrounding activity fit the legitimate distribution. A fleet of synthetic sellers may pass every individual check, but their activity, like near-identical listings and flows that consistently end at the deposit stage, along with accounts interacting mainly within the fleet, differs from the genuine population~\cite{cao2014uncovering}. A single synthetic seller in isolation may be indistinguishable from a legitimate one. A population of synthetic sellers at the scale the pipeline in Section~\ref{sec:anatomy} enables is not, because scale itself produces structure, and that structure is what action-based evaluation targets. The platform can then act accordingly: a divergent listing can be given delayed visibility, mandatory escrow, or restricted limits without any determination about the account behind it, though identity and attribution may still be needed for later audit, redress, enforcement, and liability.

\textbf{The Limits of Action-Based Evaluation.} Action-based evaluation is not a universal solution, and the paradigm we propose does not treat it as one. It works on the premise that exploitation leaves structural traces at the system level that genuine activity does not, and that these traces are detectable even when individual actors cannot be reliably distinguished. This premise holds for a wide range of threats enabled by the pipeline in Section~\ref{sec:anatomy}, including synthetic account fleets, coordinated fraud operations, and extraction-oriented workflows that deviate from genuine transaction patterns. In each of these cases, the activity produced by agentic systems at scale exhibits statistical properties that diverge from the legitimate distribution, and those properties are what action-based evaluation targets.

The premise is weaker, however, in the case of attacks that operate within the statistical envelope of genuine activity. The Infinite Impostor described in Section~\ref{sec:infinite-impostor} is the clearest example. When an agent inserts itself as a hidden intermediary between two real individuals who share an existing trust relationship, the action being performed, including the exchange of messages, the negotiation of terms, the progression of a transaction, is not structurally distinct from legitimate activity. It cannot be, because it is legitimate activity, only with an unseen party interposed. The aggregate signature of such attacks may nevertheless remain close to legitimate activity because the agent generates ordinary-looking conversations and transaction progressions around real users, real listings, and plausible objectives. The exploitation lies less in an anomalous action sequence than in the false binding between each apparent identity and the agent-controlled interlocutor presenting it. Action-based evaluation may therefore have little to target when each generated interaction remains within the statistical envelope of legitimate activity.

This is not a failure of action-based evaluation but a limit of it. The paradigm we propose does not claim that security decisions can always be made on the basis of the action alone. It claims that the center of gravity of security decisions should shift from identity to action, and that action-based evaluation should become the primary rather than supplementary mechanism. The cases where action-based evaluation is insufficient do not reverse this shift. They indicate where additional architectural mechanisms are required, mechanisms that do not attempt to detect the exploitation in flight but that constrain the conditions under which exploitation can succeed regardless of whether it is detected. The design of such mechanisms is the subject of Section~\ref{unverifiable-design}.

\subsection{Designing for the Unverifiable}\label{unverifiable-design}
The action-based paradigm of Section~\ref{identity-action} works only where exploitation leaves structural traces in the action itself, and the Infinite Impostor marks its boundary: when the agent fabricates the interlocutor while the interaction stays ordinary, no distributional signature separates it from legitimate activity. What is needed there is not better evaluation but a different defense, one that bounds the damage any single interaction can cause regardless of whether exploitation is detected. We call this bounded-damage architecture.

The economic logic here connects directly to Section~\ref{intro}: trust-based security rested on high-fidelity deception being costly enough to deter most attempts and expose the rest, and agentic automation has collapsed that cost with no path to restoring it. What remains is a different placement of cost. Generative systems cheapen persuasive content, synthetic identities, and adaptive interaction, but not the operational substrates through which attacks are delivered and monetized, whose pricing, registration, and abuse controls still shape attacker behavior~\cite{nosyk2025roots}. These substrates are persistent choke-points where cost can still be imposed. Bounded-damage architecture applies the same principle to consequential actions: where cost can no longer be imposed on fabricating a persona, it is imposed on the interaction itself, at points of settlement, privilege transfer, and externally confirmable completion that no improvement in generative fidelity can bypass. Every participant, human or synthetic, pays the same price in time, capital at risk, accumulated behavior, or delay, proportional not to who they are but to what the action can do. This does not restore the old equilibrium; it substitutes a new constraint, the structural impossibility of extracting disproportionate value from any single interaction, for the fidelity-scale tradeoff that no longer holds.

These mechanisms enforce within the environment where the interaction occurs, through the platform's control over settlement, visibility, privilege, disclosure, and execution, not through the cooperation of whoever built the agent. This is what distinguishes them from provenance and watermarking regimes, which depend on compliance at the point of generation; a platform cannot govern how every synthetic persona is produced, but it can govern the conditions under which activity on its infrastructure results in payment, access, or irreversible execution. The four mechanisms below are concrete instances, not an exhaustive catalog, each blunting a specific capability of the pipeline of Section~\ref{sec:anatomy}.

\textit{Escrow and capital-at-risk structures.} The biometric harvesting vector of Section~\ref{biometric} and the marketplace variant of the Infinite Impostor of Section~\ref{sec:infinite-impostor} both rely on the ability to convert a trusted interaction into an immediate transfer of value, whether personal data, a deposit, or an application fee. Escrow architectures interpose a platform-held settlement layer between action and extraction, releasing value only on out-of-band confirmation such as physical delivery, key handover, or verification against a state registry~\cite{park2016understanding}. A distributed ledger can encode and enforce release conditions but cannot itself establish whether an off-chain event such as a delivery, a key handover, or a valid registry entry actually occurred; that requires an oracle or other external confirmation. The security property therefore lies in separating persuasive communication from the evidence required for settlement, not in the ledger alone. The mechanism does not ask whether the seller is human or synthetic. It requires only that the transaction's real-world terminus be confirmed before value is released. An agent that convincingly impersonates a landlord cannot extract a deposit by message relay alone, because the extraction is decoupled from the conversation in which the extraction was arranged.

\textit{Rate-limited trust accumulation.} The closed-loop persona generation of Section~\ref{closed-loop} produces synthetic identities that can satisfy verification checks at the moment of account creation. Rate limits decouple the privileges available to an account from the fidelity with which it passes verification. High-consequence capabilities, including large transactions, access to sensitive data, and interaction with vulnerable populations, accrue only over time and only as a function of observed behavior, and they cannot be shortcut through any verification event, however sophisticated. The pipeline's advantage of producing convincing personas at zero marginal cost is preserved; what is removed is the corresponding ability to exploit them immediately. Rate limits impose cost without establishing permanent legitimacy. A patient adversary may be willing to accumulate standing over an extended period, and a legitimate account may be compromised or redirected after high-consequence privileges have already accrued. An agent operating under a newly malicious objective can therefore inherit the benefits of trust accumulated under earlier benign activity. Rate-limited accumulation cannot prevent these cases because account age and behavioral history remain imperfect proxies for present intent. Its contribution is narrower: it raises the cost of rapidly creating privileged account fleets. Continuous action-level evaluation, exposure caps, and delayed execution remain necessary even for long-standing accounts because previously accumulated trust cannot be treated as an unconditional authorization for future actions.

\textit{Exposure caps per interaction.} The structural inversion of Section~\ref{sec:infinite-impostor} derives its power from the collapse of per-target cost, where one interaction and one thousand interactions are roughly equivalent in attacker effort. Exposure caps impose a ceiling on what can be extracted from any individual interaction, including a maximum transaction size, a maximum disclosure per session, or a maximum privilege granted in a single exchange, independent of who the participants are. The cap applies to legitimate and illegitimate interactions alike, and precisely for that reason it cannot be defeated by higher-fidelity deception. It flattens the scale advantage by ensuring that the attacker's per-unit yield remains bounded even as their volume grows.

\textit{Cooling-off periods and delayed execution.} Agentic systems operate continuously and at a pace no human operator can match. Mandatory delays on high-consequence actions reclaim the temporal dimension that automation collapses. A large transfer that executes only after a waiting period, an account permission that activates only after a confirmation window, or a configuration change that propagates only after an out-of-band acknowledgment each introduce temporal friction that cannot be reduced by any improvement in generative fidelity. The delay is not a detection mechanism. It is a window during which other signals, including those originating outside the platform, can intervene.

Prior work suggests that defensive deception need not preserve a single consistent view and that controlled inconsistency can itself impose costs on an adversary\cite{neagoe2006}. In the Infinite Impostor setting, independently administered confirmation channels may expose the agent to observations it cannot control or perfectly reconcile, including registry state, platform records, or confirmation from the intended human. Such inconsistency does not authenticate the interlocutor, but it disrupts the coherent false social context the agent must sustain and may reveal the deception before irreversible execution. It therefore complements escrow, cooling-off periods, and exposure caps rather than replacing them.

These mechanisms do not neutralize every attack enabled by the pipeline. The Infinite Impostor remains the most instructive case. Escrow and exposure caps meaningfully bound the value an Impostor can extract per intercepted relationship, and cooling-off periods create windows in which the real participants may detect the interposition through out-of-band contact. Rate-limited trust accumulation offers limited protection when exploitation appropriates trust that already exists, when a patient adversary has accumulated standing over time, or when a previously legitimate account is subsequently compromised. These cases demonstrate why accumulated trust can affect the level of scrutiny applied to an interaction but cannot remove exposure limits or other action-level constraints. We note this limitation not as a weakness of the paradigm but as a feature of its honest scope. Bounded-damage architecture does not claim to prevent exploitation. It claims only that the consequences of any individual exploitation can be structurally constrained, and that this constraint holds even when the identity of the actor cannot be established.

What these mechanisms share, beyond their economic logic, is a relocation of the security function itself. Under the trust-based paradigm, the user was the final arbiter of authenticity, trained to recognize the artifacts of cheap deception that Section~\ref{human-in-the-loop} showed are disappearing. Under bounded-damage architecture, the user is not asked to make that determination at all. The responsibility for security is located in the design of the environment rather than in the vigilance of the participant. This is not a shift in emphasis but a shift in locus: security is exercised at the layer where it can still be exercised, and platforms become, in Lessig's sense, the regulatory substrate of the interactions they host~\cite{lessig2009code}. This reassignment is what the notion of a digital social contract names. Users accept friction they did not previously face, including slower trust accumulation, mandatory delays on high-consequence actions, and limits on the value extractable from any single interaction. Platforms accept design obligations they previously externalized to user awareness, namely the responsibility for constructing trust environments in which bounded damage holds as a structural property rather than as an aspiration. Regulators accept that mandating detection, given the arguments of Section~\ref{sec:defense-fail}, is mandating the structurally impossible, and that accountability frameworks must instead reach into architecture. The contract is not a metaphor. It is the concrete reassignment of costs and obligations that a post-trust digital environment requires in order to function.

\subsection{The Proof-of-Humanity Problem}\label{proof-humanity}

An alternative response to the same diagnosis has attracted considerable investment. If verification has become unreliable because the artifacts it checks can be synthesized, it can be made reliable again by anchoring it to something that cannot be, the human body itself. We examine this path not to endorse or reject it but as the limiting case of the trust-based paradigm, because doing so clarifies what verification costs at its extreme, and why the architectural turn of Section~\ref{unverifiable-design} answers difficulties that binding identity to the body does not resolve.

The most visible instance of this approach is the World Network (formerly Worldcoin), which has deployed iris-scanning devices in multiple jurisdictions to issue a cryptographic credential attesting that the holder is a unique human being~\cite{world_whitepaper_2026}. Adjacent projects employ different biometric anchors but share the underlying logic. A recent and intellectually serious treatment is provided by Adler et al.~\cite{adler2024personhood}, who propose ``personhood credentials'' as a privacy-preserving instrument allowing an individual to demonstrate human status to an online service without disclosing further identifying information, coordinated through zero-knowledge proofs and a network of issuers. The sophistication of this proposal deserves acknowledgment: its authors are aware of the surveillance and centralization concerns long raised against biometric registries~\cite{douceur2002sybil}, and the system they describe is designed to mitigate them. Our argument is not that such proposals are naive, but that three features of the approach are structural rather than contingent, and therefore not dissolvable by better engineering.

The first feature is centralization. A proof-of-humanity credential is useful only if it can assert global uniqueness, since the problem it is meant to solve is precisely the Sybil problem of one entity presenting as many~\cite{douceur2002sybil}. Global uniqueness requires a single authoritative registry, or a federation of registries that share sufficient information to detect cross-registry duplication. In the latter case, what is shared must be enough to establish identity, which is the very property the system was meant to abstract over. Centralization is therefore not an implementation choice made by Worldcoin or any particular vendor; it is entailed by the claim the system is attempting to make. The regulatory response confirms this structural reading: data protection authorities and courts in multiple jurisdictions have issued bans, deletion orders, and investigations citing disproportionate collection of sensitive personal data~\cite{bitpinas2025worldcoinbans}. These actions are not objections to specific operational failures that could be remedied through compliance. They are objections to the concentration of biometric authority that the approach requires to function at all.

The second feature is exclusion. Every biometric system has a non-zero false-rejection rate, and the populations for whom the rate is highest are well documented as individuals who cannot physically provide the requested biometric, those whose biometric samples are difficult to acquire reliably, and those whose cultural or religious commitments preclude the required form of data capture~\cite{worldbank_id4d_biometric_data}. Under a trust-based system in which biometrics supplement other credentials, these populations can be accommodated through alternative verification paths. Under a proof-of-humanity layer that gates digital participation, the falsely rejected are structurally excluded from whatever portion of digital life the layer covers. This is not a distribution problem solvable by expanding access or improving hardware. It is the cost of making participation conditional on a binary threshold that any biometric system will, at some rate, misapply. The exclusion scales with the ambition of the system: the more universally a proof-of-humanity credential is required, the more completely its failures translate into civic exclusion.

The third feature is surveillance. A system that can verify ``this is a unique human'' at scale establishes infrastructure capable of linking that human's actions across contexts. Zero-knowledge constructions can reduce the information disclosed at each verification event, and the Adler et al. proposal devotes considerable attention to this~\cite{adler2024personhood}. What these constructions cannot reduce is the fact that the infrastructure exists, is operated by identifiable entities, and can be repurposed. Policies restricting repurposing are reversible through legislation, regulatory capture, or state action; the infrastructure, once deployed at scale, is not~\cite{schneier2012liars}. The surveillance concern is therefore not that any particular operator will abuse the system, but that the system's existence creates an affordance whose future use is not determined by its present design.

Beneath these features lies a categorical shift. Under the prior arrangement of digital life, being human was a background presumption: individual claims, this account, this transaction, this credential, were checked against it, but the presumption itself was not demanded of each participant at each system. A proof-of-humanity layer inverts this. Humanity becomes an affirmative claim to be demonstrated at every system encountered, with the absence of demonstration becoming grounds for denial. Whether that is acceptable is a question this work does not answer, but it is one these proposals make unavoidable, and it is not reducible to how biometric data is stored or anonymized.

What emerges from these observations is not a rejection of proof-of-humanity as a research direction, but a recognition of the tension it introduces. The architectural paradigm of Section~\ref{unverifiable-design} bounds damage without requiring that identity be verified, accepting in exchange that exploitation is not prevented but only constrained. Proof-of-humanity attempts to prevent exploitation by making verification maximally reliable, accepting in exchange costs in centralization, exclusion, and the redefinition of digital participation that regulators have increasingly judged disproportionate. Each response resolves one horn of the problem by yielding on the other.

%% file: FigTex/FigPostTrust.tex
\begin{tikzpicture}
    \useasboundingbox (0,0.5) rectangle (8.6,10.0);

    \fill[dangerRed!8,rounded corners=6pt]  (0.15, 7.65) rectangle (8.45, 9.65);
    \fill[safeGreen!10,rounded corners=6pt] (0.15, -0.1) rectangle (8.45, 4.15);


    \node[font=\sffamily\bfseries\footnotesize,text=dangerRed,anchor=west] at (0.3,9.35)
        {\faLock\ \ Trust-Based Security};
    \node[font=\sffamily\scriptsize\itshape,text=dangerRed!80!black,anchor=east]
        at (8.3,9.35) {who are you?};
    \draw[dangerRed!40,line width=0.3mm] (0.3,9.15) -- (8.3,9.15);

    \foreach \i/\x/\ic/\lab in {
        1/1.55/faFingerprint/{Authenticate \\ the actor},
        2/4.30/faIdCard/{Establish \\ identity},
        3/7.05/faDoorOpen/{Permit \\ the action}
    }{
        \begin{scope}[shift={(\x,8.35)}]
            \draw[rounded corners=5pt,draw=dangerRed,line width=0.45mm,
                  top color=dangerRed!3,bottom color=dangerRed!35,
                  drop shadow={opacity=0.28,shadow xshift=1pt,shadow yshift=-1pt}]
                (-1.1,-0.55) rectangle (1.1,0.55);
            \node[text=dangerRedDark] at (-0.75,0) {\small\csname\ic\endcsname};
            \node[font=\sffamily\bfseries\scriptsize,text=dangerRedDark,
                  align=center,text width=1.5cm] at (0.25,0) {\lab};
            \node[chip=dangerRed,fill=dangerRed,text=white] at (-0.95,0.55) {\i};
        \end{scope}
    }
    \draw[flow,dangerRed] (2.65,8.35) -- (3.20,8.35);
    \draw[flow,dangerRed] (5.40,8.35) -- (5.95,8.35);

    \node[text=dangerRed,font=\small,align=center] at (7.05,7.45) {\faExclamationTriangle};
    \node[font=\sffamily\scriptsize\itshape,text=dangerRed,align=center,
          text width=3.0cm] at (7.05, 6.95) {verification is no longer reliable};

    \begin{scope}[shift={(4.3,6.0)}]
        \foreach \r/\op in {1.5/0.05, 1.2/0.1, 0.9/0.18}
            \fill[aiPurple,opacity=\op] (0,0) circle (\r);
        \draw[draw=aiPurpleDark,line width=0.6mm,rounded corners=6pt,
              left color=aiPurpleDark,right color=warnOrange]
            (-2.4,-0.55) rectangle (2.4,0.55);
        \node[text=sunYellow,anchor=west] at (-2.2,0) {\normalsize\faSync};
        \node[font=\sffamily\bfseries\footnotesize,text=white,align=center]
            at (0.15,0.10) {Paradigm Shift};
        \node[font=\sffamily\scriptsize\itshape,text=white!95,align=center]
            at (0.15,-0.22) {identity is degraded as evidence};
    \end{scope}
    \draw[flow,aiPurple,line width=0.7mm] (4.3,7.75) -- (4.3,6.65);
    \draw[flow,aiPurple,line width=0.7mm] (4.3,5.35) -- (4.3,4.30);

    \node[font=\sffamily\bfseries\footnotesize,text=safeGreenDark,anchor=west] at (0.3,3.85)
        {\faUserShield\ \ Suspect-by-Default Security};
    \node[font=\sffamily\scriptsize\itshape,text=safeGreenDark,anchor=east]
        at (8.3,3.85) {what is happening?};
    \draw[safeGreen!60,line width=0.3mm] (0.3,3.65) -- (8.3,3.65);

    \foreach \i/\x/\y/\ic/\lab in {
        1/2.25/2.75/faEye/{Evaluate \\ the action},
        2/6.35/2.75/faBalanceScale/{Measure risk \\ \& context},
        3/6.35/1.35/faHourglassHalf/{Apply friction: \\ humanity verification},
        4/2.25/1.35/faMinusCircle/{Bound damage \\ even if synthetic}
    }{
        \begin{scope}[shift={(\x,\y)}]
            \draw[rounded corners=5pt,draw=safeGreenDark,line width=0.45mm,
                  top color=safeGreen!5,bottom color=safeGreen!40,
                  drop shadow={opacity=0.28,shadow xshift=1pt,shadow yshift=-1pt}]
                (-1.7,-0.55) rectangle (1.7,0.55);
            \node[text=safeGreenDark] at (-1.35,0) {\small\csname\ic\endcsname};
            \node[font=\sffamily\bfseries\scriptsize,text=safeGreenDark,
                  align=center,text width=2.4cm] at (0.15,0) {\lab};
            \node[chip=safeGreen,fill=safeGreen,text=white] at (-1.55,0.55) {\i};
        \end{scope}
    }
    \draw[flow,safeGreen!70!black] (3.95,2.75) -- (4.65,2.75);
    \draw[flow,safeGreen!70!black] (6.35,2.20) -- (6.35,1.90);
    \draw[flow,safeGreen!70!black] (4.65,1.35) -- (3.95,1.35);

    \node[font=\sffamily\bfseries\scriptsize,text=white,align=center,
          top color=safeGreen,bottom color=safeGreenDark,
          draw=safeGreenDark,line width=0.3mm,
          rounded corners=3pt,inner sep=5pt,text width=6.5cm]
        at (4.3,0.40) {\faCheckCircle\ security decisions without resolving identity};

\end{tikzpicture}

%% file: Sections/Open.tex
\section{Tensions and Open Questions}\label{sec:open-questions}
The paradigm this work advanced does not dissolve the problem it responds to. It relocates the security function from the verification of identity to the architecture of interaction, and in doing so it inherits a set of tensions that the trust-based paradigm was able to ignore precisely because it placed the burden of authenticity on the actor. This section names the tensions we find most pressing, however we do not resolve them. We offer them as the honest accounting of what a post-trust paradigm costs, what it cannot yet do, and what remains to be worked out before the shift we propose can be more than a conceptual reorientation.

\textbf{The residual Infinite Impostor.} The Infinite Impostor forces this paradigm shift and, for the same reason, remains its hardest case. The agent need not defeat channel authentication; it runs two agent-controlled interactions in which each party sees a convincing representation of the other, so channels and accounts may authenticate correctly while the binding between account, persona, and intended human stays false. This is what makes action-level evaluation (Section~\ref{identity-action}) and bounded-damage architecture (Section~\ref{unverifiable-design}) necessary rather than optional. Within that architecture, escrow, exposure caps, and cooling-off periods bound the value extractable from each appropriated relationship and open windows for independent confirmation, but they need not reveal the false mediation itself. Whether additional mechanisms can reliably verify the binding between an apparent interlocutor and the intended person, or whether some residual risk must instead be accepted and bounded in damage, remains an open question.

\textbf{The bootstrap and measurement problem.} Action-based evaluation as proposed in Section~\ref{identity-action} rests on the availability of a baseline: a distributional model of legitimate aggregate activity against which structural divergence can be identified. This baseline is not given as it must be constructed from observed activity on the system, and the system is the very environment in which synthetic actors now operate at scale. If the observed population is meaningfully infiltrated at the moment the baseline is drawn, the baseline itself encodes the signatures of exploitation, and divergence from it ceases to be a meaningful security signal. This is not a transient difficulty that resolves once better data is collected, because there is no uncontaminated moment to collect from. A related problem attaches to evaluation. The success of bounded-damage architecture manifests as attacks that did not scale, extractions that did not complete, and relationships that were not compromised to the degree they could have been. These are counterfactual outcomes, and counterfactuals are notoriously difficult to measure. A paradigm whose success is defined by what does not happen places unusual demands on the methodologies by which it would be validated, and we do not yet know what those methodologies should look like. We regard the development of these methodologies, and the empirical evaluation of the mechanisms of Section~\ref{unverifiable-design} against them, as the next step in this research agenda.

\textbf{The normalizing pressure of the baseline.} The bootstrap problem concerns whether a baseline of legitimate activity can be measured; a distinct tension concerns what the baseline enforces even when it is measured correctly. Action-based evaluation renders activity suspect in proportion to its statistical divergence from the aggregate, and divergence is not illegitimacy. Genuine users whose behavior is atypical, whether by circumstance, culture, or idiosyncrasy, will bear a disproportionate share of the friction the architecture imposes, and users who learn this will adapt their conduct toward the distributional center. The baseline does not merely describe legitimate activity but exerts a normalizing pressure on it. Normality is not a neutral fact awaiting measurement: what registers as ordinary is historically specific and shaped by the population and power relations from which the distribution is drawn. A paradigm that makes divergence costly therefore risks consolidating the normalities it inherits. How distributional evaluation can be designed to tolerate legitimate atypicality, and who decides what range of conduct the baseline should accommodate, are among the questions this work leaves open.

\textbf{The cost of universal suspicion.} Bounded-damage architecture distributes friction across every interaction, not only the malicious ones. Escrow delays settlement for legitimate sellers as well as fraudulent ones; rate-limited trust accumulation slows genuine newcomers as well as synthetic fleets; cooling-off periods defer consequential actions for users who would never have been targeted as well as those who would. This is not incidental but structural: a mechanism that applied friction only to illegitimate interactions would be a detection mechanism, and Section~\ref{sec:defense-fail} argued at length that reliable detection is what the paradigm relinquishes. The aggregate cost of this friction is not easily estimated, but it is not zero. Airport security offers an instructive analogy, for example, a regime effective at bounding specific threats has also imposed measurable costs in time, money, and participation, and a full accounting of whether the trade was worthwhile remains contested decades after its introduction~\cite{stewart2014cost}. The question for a post-trust paradigm is whether there is a point at which the overhead of operating under universal suspicion exceeds the value digital interaction was supposed to provide. We do not know where that point is, or whether it can be known in advance rather than recognized only after it has been crossed.

\textbf{Who governs the governors?} The paradigm we propose relocates the security function from the user to the architecture of the platform, and in doing so assigns platforms a role closer to that of a regulator than of a service provider. Lessig's observation that code is law~\cite{lessig2009code} takes on a sharper edge under bounded-damage architecture, because the mechanisms that constrain exploitation (e.g., escrow thresholds, rate limits on trust accumulation, exposure caps, mandatory delays) are not incidental product decisions but the substantive rules under which digital interaction proceeds. Trust-based security could at least maintain the fiction that platforms were neutral substrates on which verified actors transacted; the paradigm we describe makes the platform's regulatory role explicit and unavoidable, which raises the question of what governs the platforms in turn.

Adoption turns on which costs of deception a platform internalizes. When declining trust erodes the participation and revenue a platform depends on, as the Akerlof dynamic of Section~\ref{sec:reputationEconomics} implies, the incentive to deploy escrow, settlement delays, and exposure limits is direct. Payment and marketplace systems already do so when fraud threatens settlement integrity. But where those costs fall on users, counterparties, financial institutions, or public authorities, a platform may rationally underinvest. The architecture thus poses a distributional question as much as a technical one: who bears the cost of precaution, and who bears the loss when it is absent? Voluntary adoption is most plausible where the platform internalizes both sides; elsewhere, regulatory obligation may be required.

Legal accountability presents a related difficulty. Liability frameworks developed around the hosting and moderation of content do not necessarily resolve responsibility for systems that structure settlement, privilege allocation, escrow, or irreversible execution. The relevant governance question is therefore not only whether a platform knew that particular content was deceptive, but whether an intermediary exercising architectural control took reasonable steps to constrain foreseeable transactional harm. The appropriate allocation of such duties remains jurisdiction-dependent and unresolved, but the distinction between hosting expression and designing the conditions of execution is central to the regulatory problem posed by bounded-damage architecture.

Trust-based accountability mechanisms lose much of their purchase against an architecture whose security function is constitutive rather than procedural. Mandating specific escrow windows or exposure caps pushes regulators into system design rather than review, a role for which the institutional machinery does not yet exist. This concentration of architectural power is not a side effect but a requirement of the paradigm. Whether democratic societies can build adequate accountability frameworks, and what they would look like, this work leaves open.

\textbf{Three futures.} A paradigm argument is evaluated by history rather than by experiment, and it is worth stating, hypothetically, what the candidate histories look like. In the first, the paradigm materializes: bounded-damage mechanisms become default properties of platform design, and the question of who is acting recedes from the center of security decisions. Success of this kind would not dissolve the tensions named above, but instead would promote them to the live problems of a suspect-by-default world. In the second, the shift is preempted by a different change in how security is done: provenance and attestation infrastructure matures, personhood credentials succeed at acceptable cost despite Section~\ref{proof-humanity}, or regulation of the small number of platforms on which agentic capability is concentrated constrains misuse at its source. Each route relocates trust rather than dispensing with it; in this future the paradigm is not refuted so much as never required. In the third, the paradigm never takes hold: platforms are penalized for unilaterally adopting friction, the losses of deception are externalized, and regulators mandate detection because it is auditable even where it is futile~\cite{nemecek2025watermarking}. We offer these futures not as predictions but as the terms in which the viability of a post-trust paradigm can be debated.

%% file: Sections/Conclusion.tex
\section{Conclusion}
\label{conclusion}
We have argued that a particular paradigm of security is exhausted, one that treated the fidelity-scale tradeoff as a stable constraint and built its detection systems, verification mechanisms, and user training around the artifacts that cheap deception reliably produced. Agentic AI does not weaken this paradigm; it removes the condition under which it was coherent. What follows from that diagnosis is not the end of security as a practice, but the end of an approach to security that has so thoroughly shaped the field that its assumptions have become invisible. The suspect-by-default paradigm, with its shift from identity to action and its turn toward bounded-damage architecture, is not offered as a solution in the sense that the exhausted paradigm understood that word. It does not restore the distinguishability between authentic and synthetic that generative progress has eliminated, and it does not promise that exploitation can be prevented. It proposes instead that security be exercised over what actions a system permits, what value any interaction can extract, and what constraints hold regardless of who is on the other end. Whether that is still ``security'' in the field's historical sense we leave open: it may be better named a discipline of bounded harm, or it may be that security was always the practice of making exploitation costly, here pursued where the cost must fall somewhere other than on the attacker. The Infinite Impostor, the bootstrap and measurement problem, the normalizing pressure of the baseline, the cost of universal suspicion, and the question of who governs the governors are not the tensions of a field in crisis but of a field in transition.
